\documentclass[twocolumn]{aa} 

\usepackage[dvipsnames]{xcolor}

\usepackage{tabularx}
\usepackage{array}

\newcolumntype{Y}{>{\centering\arraybackslash}X}

\usepackage{graphicx}
\usepackage{txfonts}
\usepackage{hyperref}

\begin{document} 

\newcommand{\afffias}{Frankfurt Institute for Advanced Studies (FIAS), Ruth-Moufang-Strasse~1, 60438 Frankfurt am Main, Germany}
\newcommand{\affbgu}{Physics Department, Ben-Gurion University of the Negev, Beer-Sheva 84105, Israel}
\newcommand{\affbul}{Institute for Nuclear Research and Nuclear Energy, Bulgarian Academy of Sciences, Sofia, Bulgaria}

\title{Dark Energy as a critical period in Binary Motion: Bounds from Multi-scale Binaries}

\author{David Benisty\inst{1,2} \and  Jenny Wagner\inst{3} \and Denitsa Staicova\inst{4} }

   \institute{$^{1}\,$ Frankfurt Institute for Advanced Studies (FIAS), Ruth-Moufang-Strasse~1, 60438 Frankfurt am Main, Germany 
\\$^{2}\,$ Kavli Institute of Cosmology (KICC), University of Cambridge, Madingley Road, Cambridge, CB3 0HA, UK
\\$^{3}\,$ Bahamas Advanced Study Institute and Conferences, 4A Ocean Heights, Hill View Circle, Stella Maris, Long Island, The Bahamas
\\$^{4}\,$ Institute for Nuclear Research and Nuclear Energy, Bulgarian Academy of Sciences, Sofia, Bulgaria}
   \date{}

 
\abstract{The two-body problem under the influence of both dark energy and post-Newtonian modifications is studied. In this unified framework, we demonstrate that dark energy plays the role of a critical period with $T_{\Lambda} = 2\pi/c \sqrt{\Lambda} \approx 60~\text{Gyr}$. We also show that the ratio between orbital and critical period naturally emerges from the Kretschmann scalar, which is a quadratic curvature invariant characterizing all binary systems effectively represented by a de Sitter-Schwarzschild spacetime. The suitability of a binary system to constrain dark energy is determined by the ratio between its Keplerian orbital period $T_\text{K}$ and the critical period $T_\Lambda$. Systems with $T_\text{K} \approx T_\Lambda$ are optimal for constraining the cosmological constant $\Lambda$, such as the Local Group and the Virgo Cluster. Systems with $T_{\text{K}} \ll T_\Lambda$ are dominated by attractive gravity (which are best suited for studying modified gravity corrections). Systems with $T_{\text{K}} \gg T_\Lambda$ are dominated by repulsive dark energy and can thus be used to constrain $\Lambda$ from below. 
We use our unified framework of post-Newtonian and dark-energy modifications to calculate the precession of bounded and unbounded astrophysical systems and infer constraints on $\Lambda$ from them. Pulsars, the solar system, S stars around Sgr A*, the Local Group, and the Virgo Cluster, having orbital periods of days to gigayears, are analyzed. {\it The results reveal that the upper bound on the cosmological constant decreases when the orbital period of the system increases, emphasizing that $\Lambda$ is a critical period in binary motion.}}
  


%

\maketitle

\section{Introduction}
The surprising discovery that the cosmic expansion is accelerating led to propose the existence of a mysterious dark energy that makes up 68\% of the energy of the observable universe \cite{Perlmutter_SN,Peebles:2002gy}. The simplest model for dark energy, the cosmological constant $\Lambda$, assumes a constant energy density that accelerates the galaxies further away from us. Signatures of dark energy emerge from different probes: in the late universe from supernovae~\cite{Riess_SN, Perlmutter_SN} and Baryon Acoustic Oscillations~\cite{Cole_BAO, Eisenstein_BAO,Adil:2021zxp}, in the early universe from the Cosmic Microwave Background (CMB) radiation~\cite{Planck:2018vyg}. All of these probes rely on measurements of the cosmic expansion rate detected at large distances. However, the effect of dark energy could also occur in observations at much lower distances, as recently found~\cite{Benisty:2023vbz} and numerous other studies discuss the impact of dark energy in the local universe ~\cite{Chernin:2000pq,Baryshev:2000kw,Chernin:2001nu,Kim:2020gai,Karachentsev:2003eh,Chernin:2003qd,Teerikorpi:2005zh,Chernin:2009ms,Chernin:2006dy,Peirani:2008qs,Chernin:2010jt,Teerikorpi:2010zz,Chernin:2015nna,Chernin:2015nga,Silbergleit:2019oyx}. The cosmological constant changes the predicted mass of the Local Group (LG), assuming it mainly consists of the Milky Way and Andromeda forming a two-body system~\citep{Carlesi:2016eud,Gonzalez:2013pqa,Li:2007eg,vanderMarel:2012xp,Chernin:2009ms,Hartl:2021aio,Phelps:2013rra,McLeod:2016bjk,Lemos:2020vhj}. Therefore, it is important that the contribution of dark energy should be taken into account even at such short distances~\cite{Penarrubia:2014oda}. 

\begin{figure}
    \centering
\includegraphics[width=0.24\textwidth]{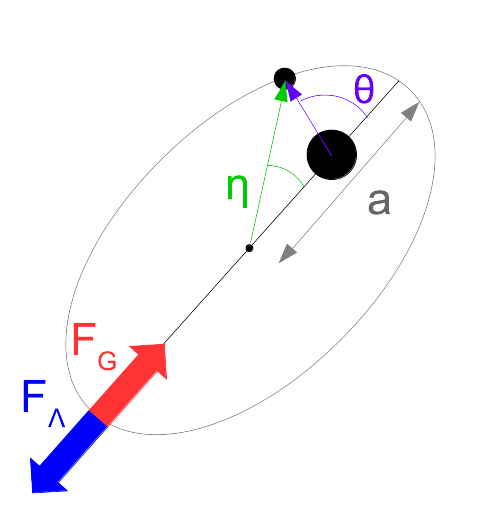}
\caption{Interaction between two bodies in an effective central potential as considered in this paper: gravitational force (red arrow) as an attracting antagonist against the repulsive force from the cosmological constant (blue arrow). The eccentric anomaly is denoted by $\eta$, the true anomaly by $\theta$, the semi-major axis of the orbit is called $a$.}
    \label{fig:notation}
\end{figure}

In an isolated two-body system, as shown in Fig.~\ref{fig:notation}, $\Lambda$ changes the radial force by a modification with linear dependence of the separation between the two bodies:
\begin{equation}
\ddot{r} = - G\frac{M}{r^2} + \frac{1}{3}\Lambda c^2 r \;,
\label{eq:eom}
\end{equation}
where $r$ is the separation, $M$ is the total mass of the binary system, $G$ is the Newtonian gravitational constant, and $c$ is the speed of light. It is then possible to quantify the impact of $\Lambda$ versus the Newtonian force via the ratio between the two terms on the right-hand side of Eq.~(\ref{eq:eom}):
\begin{equation}
\kappa \equiv \left(\frac{T_{\text{K}}}{T_\Lambda}\right)^2 \;, 
\label{eq:kappa}
\end{equation}
where the Keplerian period, depending on the semi-major axis of the orbit $a$, is given by 
\begin{equation}
T_{\text{K}} = 2\pi \sqrt{\frac{a^3}{GM}} 
\label{eq:Tkep}
\end{equation}
and the critical period related to the cosmological constant is 
\begin{equation}
T_{\Lambda} = \frac{2\pi}{c \sqrt{\Lambda}} \;.  
\label{eq:TLambda}
\end{equation}
If $\kappa < 1$ (or $T_{\text{K}} < T_{\Lambda}$), the Newtonian force prevails. If $\kappa > 1$ (or $T_{\text{K}} > T_{\Lambda}$), the repulsion caused by the cosmological constant dominates.

This condition has been found in \cite{Benisty:2023vbz} where it was shown that, for Milky Way and Andromeda as a binary system, $\kappa \approx 10\%$. This implies a dominating Newtonian force for this system on the one hand. But, on the other, dark energy still pulls Andromeda apart. Since the velocities are in the Newtonian regime, \cite{Benisty:2023vbz} does not consider the post-Newtonian (PN) correction to the solution for this binary motion.

In this paper, we extend the approach of \cite{Benisty:2023vbz} and solve the equations of motion for general binary systems including both, the effect of a cosmological-constant term (called ''post-cosmological'', PC), and the PN terms. Thus, contrary to existing approaches which mostly focus either on the limit of vanishing cosmological contribution to test alternative theories of gravity or vice versa, our unified approach can produce the orbital characteristics for both limits, as well as the transition regime in between. Applying this unified formalism to different bounded systems, including, among others, those considered in \cite{Je06,Baker:2014zba,Brax:2018bow}, we find that the upper bound on $\Lambda$ is closer to the value inferred from the CMB, when the period of the system is close to $T_\Lambda$. 

This paper also discusses the fundamental reason, based on differential geometry, why $T_\Lambda$ as the critical period is the relevant scale for the impact of dark energy on two-body motions. Often, in the literature, one finds the vague argument that $\Lambda$ is only relevant on Gpc scales, implying that a characteristic \emph{length} scale mattered, when, actually, as we show here, it is the \emph{time} scale of the orbital period that defines the significance of the dark energy term in binary motion.

The structure of the paper is as follows: Section~\ref{sec:astro} introduces different astrophysical systems and compares the estimated impact of the cosmological constant on their motion to highlight suitable probes which can put the strongest constraints on $\Lambda$ on small scales. Section~\ref{sec:action} derives the analytical solution of the equations of motion in the presence of both, dark energy and the first PN correction. Section~\ref{sec:upper} calculates the constraints on dark energy from several binary systems introduced in Section~\ref{sec:astro} for which the required observational data is available. Subsequently, Section~\ref{sec:dis} summarizes the results.

\begin{figure*}
    \centering
\includegraphics[width=0.8\textwidth]{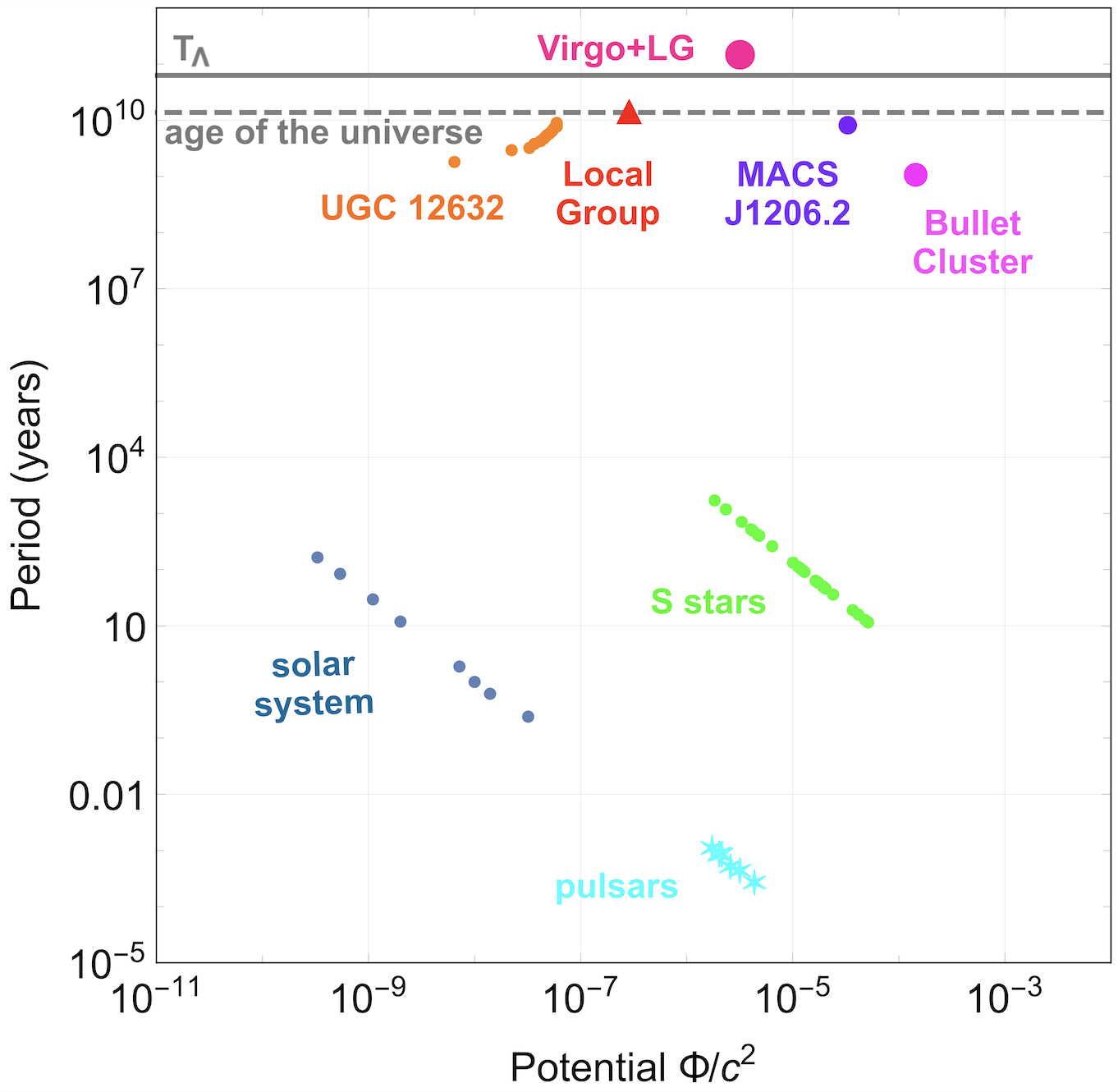}
\caption{Scaled Newtonian gravitational potential $\beta = G M/a c^2$, versus the measured and estimated orbital periods for various astrophysical systems. The critical period $T_{\Lambda} = 2\pi/c \sqrt{\Lambda}\approx 60~\text{Gyr}$ (Eq.~\ref{eq:TLambda}) sets the scale above which dark energy dominates over gravity. For systems with $\beta \preceq 1$ the post-Newtonian correction to General Relativistic effects can be observed.}
    \label{fig:periods}
\end{figure*}

\section{Astrophysical Binary Systems}
\label{sec:astro}

Many astrophysical systems across various size and mass scales can be modelled as binary systems and can be used as probes to constrain theories of gravity and the impact of dark energy.
To evaluate the suitability of a binary system for such studies two characteristic quantities have to be calculated. The first is the Newtonian gravitational potential $\Phi$, which we scale as
\begin{equation}
\beta = \frac{\Phi}{c^2} = \frac{G M}{r c^2} \;,
\label{eq:beta}
\end{equation}
typically inserting $r=a$, the semi-major axis of the orbit.
The second quantity is the Keplerian period $T_\text{K}$, as defined in Eq.~(\ref{eq:Tkep}). Then, the distinction between different regimes can be made based on the ratios between $T_\text{K}$ with respect to $T_\Lambda$ and $\beta$ to 1. 

\subsection{Binary systems considered}

In Fig.~\ref{fig:periods}, we plot the following astrophysical systems that can be approximated as binary systems with different $\beta$ and $\kappa$:

\begin{itemize}
\item \textbf{Solar System} -- 
Since these systems have periods of the order of years or tens of years, which is much smaller than $T_{\Lambda}$ ($T_\text{K} \approx 1~\text{yr} \approx 10^{-11} \, T_{\Lambda}$), the effect of dark energy in these systems is negligible. Ref. \cite{Kagramanova:2006ax} constrains dark energy from the solar system, and obtains an upper bound of $\Lambda < 10^{-37}\, \text{m}^{-2}$. 
\item \textbf{S Stars} -- The centre of the Milky Way hosts the closest supermassive black hole, Sgr A*. The stars orbiting Sgr A* are called S stars with decades of monitoring of their locations and velocities \cite{Yu:2016nzn,Abuter:2018drb,Do:2019txf,Abuter:2020dou,Amorim:2019hwp,Dialektopoulos:2018iph,Borka:2021omc,Capozziello:2014rva,Capozziello:2014mea,Borka-Jovanovic:2019mya,Will:2018ont,Will:1997bb,Scharre:2001hn,Moffat:2005si,Zhao:2005zq,Bailey:2006fd,Deng:2009tg,Barausse:2012da,Borka:2012tj,Enqvist:2013tsa,Borka:2013dba,Capozziello:2014rva,Berti:2015itd,Borka:2015vqa,Zakharov:2016lzv,Zhang:2017srh,Dirkes:2017ecu,Pittordis:2017byg,Hou:2017cjy,Nakamura:2018yaw,Banik:2018ydl,Dialektopoulos:2018iph,Kalita:2018ubo,Will:2018ont,Banik:2019zme,Pittordis:2019kxq,Nunes:2019bjq,Anderson:2019eay,Gainutdinov:2020bbv,Bahamonde:2020bbc,Banerjee:2020rrd,Ruggiero:2020yoq,Okcu:2021oke,deMartino:2021daj,DellaMonica:2021xcf,DAddio:2021xsu,2017ApJ...837...30G,Davis:2023zqv}. A large fraction of these stars have orbits with high eccentricities. Thus, they reach high velocities at the pericentre and can be used for constraining scalar interactions. The period of the stars varies from about 10 to about 1000~years, being circa $10^{-10} \, T_{\Lambda}$ to $10^{-8} \, T_{\Lambda}$. Therefore, the effect of dark energy in these systems is negligible. The precise bound of dark energy from S stars will be explored in Section~\ref{sec:upper}. 
\item \textbf{Binary Pulsars} -- The first evidence for gravitational waves was provided by the binary Hulse-Taylor pulsar PSR~B1913+16 \citep{1981SciAm.245d..74W,Weisberg:2016jye}. 
Pulsar systems consist of a neutron star with a white dwarf or another neutron star companion. The monitoring of the times of arrival of the pulsar's radio pulses allows us to infer the properties of the orbit, such as the precession and the emission rate of gravitational waves \citep{1986AIHS...44..263D}.
Therefore, they are useful tools for testing gravity due to the extreme precision of the radio pulses they emit. For instance, post-Keplerian parameters which take different forms in different theories of gravity have been constrained from pulsar measurements \cite{Liu:2017xef,DeLaurentis:2011tp,DeLaurentis:2013zv,DeLaurentis:2013fra,DeLaurentis:2013zv,Dyadina:2016bzb,Narang:2022jkv,Brax:2019tcy,Davis:2019ltc,Brax:2019tcy,Brax:2020vgg,Brax:2021qqo,Benisty:2021cmq,Benisty:2022lox}.
The period of pulsars is about $10^{-3}~\text{yr} \approx 10^{-14} \, T_{\Lambda}$. Their constraint on dark energy is explored in~\cite{Je06}. In Section~\ref{sec:upper}, we give the latest update on this constraint with respect to the pulsars' period.  
\item \textbf{Local Group} -- 
Modelling the LG as a binary system of its two largest member galaxies, Milky Way and Andromeda, the binary motion of these two galaxies forms a bounded system with $T_\text{K} \approx 17~\text{Gyr} \approx 0.27 \, T_{\Lambda}$. Thus, as detailed in \cite{Benisty:2023vbz}, the system can be used to infer an upper bound on $\Lambda$ after biases due to small-mass perturbers and the embedding environment have been accounted for. Here, PN corrections are negligible, so that the system cannot serve as a probe for modified gravity tests.
\item \textbf{Bullet Cluster (1E 0657--558)} -- 
As noted in \cite{Bisnovatyi-Kogan:2019iyz}, binary systems on galaxy-cluster scale are promising probes of $\Lambda$, which is why we consider the two receding mass clumps, ``bullets'', in the galaxy cluster 1E 0657--558. Using the distance between the clumps from \cite{Clowe} and the total mass of the system from \cite{Lage}, we estimate $T_\text{K} \approx 1.23~\text{Gyr} \approx 0.02\, T_\Lambda$, such that its suitability to constrain $\Lambda$ is worse than the LG but still worth investigating.
%
\item \textbf{Galaxy Cluster MACS J1206.2} -- 
To compare the suitability of an unrelaxed cluster collision, like the Bullet Cluster, with the rotation curves of a relaxed cluster, we estimate the out-most bound orbit around the centre of MACS J1206.2 to be at its virial radius. Together with the virial mass from \cite{Umetsu}, we arrive at $T_\text{K} \approx 8.2~\text{Gyr} \approx 0.14 \, T_\Lambda$. Thus, galaxy-cluster rotation curves could be as promising as LG-like systems to constrain $\Lambda$. 
%
\item \textbf{Virgo Cluster and LG} -- 
Considering the LG and the Virgo Cluster as a binary system and neglecting the impact of the embedding environment, as done in \cite{Bisnovatyi-Kogan:2019iyz}, one finds that the system is unbounded because $T_\text{K} \approx 172~\text{Gyr} \gg T_{\Lambda}$. Hence, it is no suitable probe for theories of gravity but a promising probe for $\Lambda$. Yet, we first have to investigate that the formalism as introduced in Section~\ref{sec:action} for bounded orbits with perturbative PN and PC corrections can be applied to this system.
\end{itemize}

\subsection{Advantages and disadvantages}
For binary systems with periods very short compared to $T_\Lambda$ and observing times, like for instance, pulsars, their period can be measured many times resulting in a very high precision which is necessary to constrain PN corrections to the unperturbed Keplerian orbits in Newtonian gravity. Besides this, short periods allow us to track the orbits and infer their characteristic parameters from observables, as is the case for the planets of the solar system or the S stars around Sgr A*. Since these probes are also found at small distances away from us, they do not require to assume any cosmological background model for their interpretation, such that they are less dependent on model assumptions, for instance, about a distance-redshift relation or the presence of additional dark matter and its distribution. Therefore, these probes are ideal to constrain PN corrections.

In contrast to them, objects at higher distances, like the LG or the bullets in the Bullet Cluster are more suitable to constrain dark energy since cosmic expansion plays a non-negligible role in their motion. 
These binary systems have longer Keplerian periods which come at the cost that we only observe a tiny fraction of their period and their orbits in space. Therefore, additional models have to be adopted to solve the equations of motion. Another interesting aspect of these probes is the fact that the LG as a binary system and galaxy rotation curves (like the one for UGC~12632 shown in Fig.~\ref{fig:periods}) are almost on par in terms of their constraining quality. In contrast, rotation curves on galaxy-cluster scale seem to yield better constraints on $\Lambda$ compared to merging cluster clumps, which can be seen on Fig.~\ref{fig:periods} comparing the estimates of the Keplerian period for the galaxy cluster MACS~J1206.2 and the Bullet Cluster. The reason for this behaviour can be traced to the sizes of the semi-major axes and masses to be inserted into Eq.~(\ref{eq:Tkep}).

For the same reason, galaxy clusters are not superior to the LG in constraining $\Lambda$, as one might have thought. Even if their Keplerian periods are almost of the same order of magnitude, we have to replace the highly precise data from Milky Way and Andromeda by cosmological and other model assumptions to constrain $\Lambda$ from clusters at cosmic distances. Hence, the confidence bounds on $\Lambda$ need not be smaller than those obtained from the LG measurements. 

\subsection{Relation to the Hubble tension}

Determining $T_\Lambda$, there are two choices for $\Lambda$ to be inserted into Eq.~(\ref{eq:TLambda}) to obtain the values shown in Fig.~\ref{fig:periods}. 
From \cite{Planck:2018vyg}, a constraint on $\Omega_m$ in a flat Friedmann cosmology directly yields $\Omega_\Lambda$, which determines $\Lambda$ when inserting a value of the Hubble constant, $H_0$, into
\begin{equation}
\Lambda = 3 \Omega_\Lambda \left(\frac{H_0}{c}\right)^2 \;.
\label{eq:LambdaExpansion}
\end{equation}
Using $H_0=(67.36\pm 0.54)~\text{km}/\text{s}/\text{Mpc}$ of \cite{Planck:2018vyg}, 
the critical period is
\begin{equation}
 T_\Lambda^{\rm Planck} = (63.66 \pm 0.55)~\text{Gyr} \;. 
\end{equation}
Using $H_0=(73.6 \pm 1.1)~\text{km}/\text{s}/\text{Mpc}$ measured from the Pantheon+ and SH0ES data set instead, \cite{Brout:2022vxf}, we obtain 
\begin{equation}
T_\Lambda^{\rm SH0ES} = (59.11 \pm 1.14)~\text{Gyr} \;.  
\end{equation}
Thus inserting two different values of $H_0$ which are discrepant at the 5-$\sigma$ level, we note that their corresponding $\Lambda$ values have a $3.6$-$\sigma$ difference. As we do not focus on the $H_0$-tension in this work, we only use $H_0$ from \cite{Planck:2018vyg}. The critical period in Fig.~\ref{fig:periods} is also calculated from that value.

\section{Post-Newtonian-Post-Cosmological joint formalism}
\label{sec:action}
\subsection{Fundamental justification}
\label{sec:fundamentals}

All binary systems introduced in Section~\ref{sec:astro} can be treated by the same formalism as an effective central potential and a test body with reduced mass orbiting in this (PN) potential with or without including dark energy terms. The joint, general line element $ds^2$ for the effective central potential and the effect of dark energy reads
\begin{equation}
ds^2 = -f(r) dt^2 + \frac{dr^2}{f(r)} + r^2 \left(d\theta^2 + \sin^2(\theta)d\phi^2\right) \;,
\label{eq:le}
\end{equation}
in which $\theta$ and $\phi$ are the usual spherical angles and 
\begin{equation}
f(r) = 1 - \frac{2 G M}{c^2 r} - \frac{1}{6}\Lambda r^2 
\label{eq:f}
\end{equation}
captures the gravitational-potential and dark-energy contributions.
It is thus a de Sitter-Schwarzschild metric which reduces to the Schwarzschild metric in the limit of $\Lambda=0$. Eq.~(\ref{eq:f}) directly contains the dimensionless parameter $\beta$ (Eq.~\ref{eq:beta}) that will later be used to track PN corrections. Expressing $\beta$ in terms of the Schwarzschild radius $r_\text{s}=2GM/c^2$, it reads $r_\text{s}/(2r)$.

The role of $\kappa$ becomes clear when we determine the quadratic scalar that quantifies the gauge-invariant curvature of spacetime, i.e. the Kretschmann scalar, defined as
\begin{equation}
K \equiv R^{\lambda}_{\;\alpha\beta\gamma} R_{\lambda}^{\;\alpha\beta\gamma} \;,
\end{equation}
in which $R^{\lambda}_{\,\alpha\beta\gamma}$ is the Riemann tensor. 
For the de Sitter-Schwarzschild metric, we obtain
\begin{equation}
K_\text{SdS} = \frac{48 \,G^2 M^2}{c^4 r^6}+ \frac{2}{3} \Lambda^2 =
48\frac{\omega_{\text{K}}^4}{c^4} + \frac{2}{3} \frac{\omega_\Lambda^4}{c^4} \;,
\label{eq:K}
\end{equation}
using $r=a$ in the last step and introducing the Keplerian angular frequency $\omega_\text{K} = 2\pi/T_\text{K}$.
The curvature scalar can thus be written as a combination of the two periods $T_\text{K}$ and $T_\Lambda$.
Alternatively, we can rewrite the periods in terms of angular frequencies, such that $\kappa=(\omega_\Lambda/\omega_\text{K})^2$, to avoid introducing factors of $2\pi$ or express $K$ as
\begin{equation}
K_\text{SdS}  = K_\text{s} \left( 1 + \frac{\kappa^2}{72} \right) \;,
\label{eq:K2}
\end{equation}
in which $K_\text{s}=48 \, G^2 M^2/(c^4 r^6)$ is the Kretschmann scalar for the Schwarzschild metric. Eq.~(\ref{eq:K2}) shows that $\kappa$ is the characteristic term to describe the impact of $\Lambda$ when we embed the two-body motion, effectively modeled as orbits in a Schwarzschild metric, into an expanding de Sitter space. So, the ratio between the Keplerian orbital period to the critical period $T_\Lambda$, or the corresponding ratio of angular frequencies, arising in the solution of \cite{Benisty:2023vbz} emerges naturally from the more fundamental origin of a curvature invariant in General Relativity. This implies that $\kappa$ as a ratio of time-related quantities determines the impact of dark energy on a binary system and not a ratio of length scales, as is often given.

After deriving Eq.~(\ref{eq:K}), it is clear that our overview of probes of PC and PN corrections in Fig.~\ref{fig:periods} can be directly linked to \cite{Baker:2014zba} because this work also used $\Phi/c^2$ and the Kretschmann scalar to characterize tests of gravity on various scales. 

\subsection{Critical Frequency vs. Critical Distance}
There are different criteria to test the dominance of dark energy. One of them is based on the zero-velocity surface \cite{Kim:2020gai} while another is based on the zero-acceleration surface~\cite{Pavlidou:2013zha,Pavlidou:2014aia,Tanoglidis:2014lea}. In principle, however, all of these criteria use the same equilibrium in which the outward-directed influence of $\Lambda$ is balanced by the inward-directed gravitational attraction as reference. For instance, the zero velocity surface reads:
\begin{equation}
r_\mathrm{V}^3 = \frac{G M}{\Lambda c^2} \;,
\label{eq:critdis}
\end{equation}
which corresponds to $\kappa = 3$ in our framework. Inserting Eq.~\ref{eq:critdis} into Eq.~\ref{eq:K}, we see that the ratios of critical frequencies are equivalent to a ratio of length scales. Ref.~\cite{Benisty:2023vbz} shows that the period of the LG is about 30\% of the critical period, hence dark energy is important in this binary motion. This is equivalent to the condition that the distance between the Milky Way and Andromeda galaxies is $\sim 0.77$~Mpc, which is of the same order of magnitude as the radius of the zero-acceleration surface of about $2$~Mpc. We prefer to use the critical frequency, since the critical distance $r_\mathrm{V}$ depends on the mass of system and $\Lambda$. In contrast to this, the frequency condition separates cosmology from the system considered. It is a direct comparison between the critical one only depending on $\Lambda c^2$ and the Keplerian one only depending on the characteristics of the binary system. Thus, similar to defining a Planck time, the critical period only depends on a cosmological parameter independent of the system under consideration.

\subsection{Lagrangian and action}
As also discussed in~\cite{Damour:1988mr,Bisnovatyi-Kogan:2023aqf,Je06}, after transforming into the centre-of-mass system, the Lagrangian per reduced mass reads
\begin{equation}
\mathcal{L} = \mathcal{L}_0 + \mathcal{L}_\Lambda + \frac{1}{c^2} 
 \mathcal{L}_1^{\rm{PN}}
\end{equation}
with the three individual components
\begin{eqnarray}
&&
\mathcal{L}_0 = \frac{1}{2}v^2 + \frac{G M}{r} \;, \quad \mathcal{L}_\Lambda = \frac{1}{6} \Lambda c^2 r^2 \;,
\nonumber \\
&&
\mathcal{L}_{1}^{\rm{PN}} = \tfrac{1-3 \nu}{8}v^4 + \tfrac{G M}{2 r}
\left[(3 + \nu)v^2 +\nu (v\cdot \hat{n})^2 - \tfrac{G M}{r}\right] \;.
\nonumber
\end{eqnarray}
The action $\mathcal{L}_{1}^{\rm{PN}}$ is the first PN correction to the Newtonian term $\mathcal{L}_0$ \cite{AIHPA}. The cosmological-constant action in binary motion is $\mathcal{L}_{\Lambda}$~\cite{Carrera:2006im}. We denote the reduced mass as $\mu$ and the total mass of the system as $M$. In addition, $r$ is the separation between $M$ and $\mu$ and $v$ their relative velocity. To simplify notations, $\nu \equiv \mu/M$ is the ratio between the reduced mass and the total mass with $0 \leq \nu \leq 1/4$~\cite{AIHPA}. For a binary system whose motion occurs in the polar coordinates $(r, \theta)$, we relate $v$ and $r$ with the derivatives of $r$ and $\theta$ to obtain
\begin{equation}
\dot{r} = v\cdot \hat{n} \;, \quad v^2=\dot{r}^2+r^2 \dot{\theta}^2 \;, \quad r^2\dot{\theta} = |\vec{r}\times \vec{v}| \;.
\end{equation}
Here, $\hat{n}$ denotes the unit vector in direction of the separation and $\vec{r}$ and $\vec{v}$ are the separation and velocity vectors, respectively. Consequently, the energy per units of reduced mass is: 
\begin{equation}
\epsilon \equiv \frac{E}{\mu} = \frac{1}{2}v^2 - \frac{G M }{r} - \frac{1}{6} \Lambda c^2 r^2 + \frac{\epsilon_1}{c^2} \;,
\end{equation}
with the first PN correction
\begin{eqnarray}
\epsilon_1 = \frac{3}{8}(1-3 \nu)v^4 + \frac{G M}{2 r}\left[(3+\nu)v^2 +\nu \dot{r}^2 + \frac{G M }{r}\right] \nonumber \;.
\end{eqnarray}
The angular momentum per units of reduced mass is
\begin{equation}
\vec{j} \equiv \frac{\vec{J}}{\mu} = \left[1 + \frac{1}{c^2}\left(\frac{v^2}{2}(1-3 \nu)  + (3 + \nu)\frac{G M}{r}\right) \right] \vec{r} \times \vec{v} \;.
\end{equation}

\subsection{Effective potential}
\label{sec:effective_pot}
In order to solve the equations of motion, we isolate $\dot{r}$ and $\dot\theta$ from the conservation of the energy and the angular momentum, see Section~\ref{sec:eom} below. To do so, we write 
\begin{equation}
\dot{r}^2 = \sum_{i = -2}^{3}{\frac{\alpha_i}{r^i}} \;,\quad \dot{\theta}r^2 = \sum_{i = -2}^{2}{\frac{\gamma_i}{r^i}}
\label{eq:fullEoM}
\end{equation}
with coefficients $\alpha_i$ 
\begin{eqnarray}
\alpha_{3} &=& 
\left( 8 -3\nu \right) \frac{G M j^2}{c^2} \;, \nonumber \\
\alpha_{2} &=& 
-j^2 \left(1 - 2 (1-3 \nu ) \frac{\epsilon}{c^2}\right) + 5\left(\nu - 2\right)\left(\frac{G M}{c}\right)^2 \;, \nonumber \\
\alpha_{1} &=& 
2 G M \left(1+\frac{\epsilon}{c^2}(7 \nu -6) \right) \;, \nonumber \\
 \alpha_{0} &=& 
 2\epsilon\left[1 + \frac{3}{2}\left( \frac{\epsilon}{c^2} - \frac{j^{2}}{9\epsilon} \Lambda \right) \left(3\nu - 1\right) \right] \;,  \nonumber\\
\alpha_{-1} &=& 
G M \Lambda\frac{7\nu - 6}{3} \;, \nonumber\\
\alpha_{-2} &=& \frac{1}{3}\Lambda  c^2 \left(1 + 3 \left(3\nu -1\right) \frac{\epsilon}{c^2}\right) \;.
\label{eq:alphas}
\end{eqnarray}
The cosmological constant adds a quadratic correction to the potential and a linear term from the interaction with the PN correction.
For the second equation, we introduce
\begin{eqnarray}
\gamma_{1} &=& 2 j \left( \nu - 2\right) \frac{G M}{c^2} \;, \nonumber \\
\gamma_{0} &=& j \left(1 + (3 \nu - 1 )\frac{\epsilon}{c^2} \right) \;,\nonumber  \\
\gamma_{-2} &=& \frac{3\nu-1}{6} j \Lambda \;,
\end{eqnarray}
having $\gamma_{-1}=0$. 

The effective potential can be rescaled with respect to the Schwarzschild radius using the dimensionless quantities
\begin{equation}
x = \frac{r}{r_\mathrm{s}} \;, \quad \tilde{\epsilon} = \frac{\epsilon}{c^2} \;, \quad \tilde{j} = \frac{j}{r_\mathrm{s} c} \;, \quad \tilde{\Lambda} = \Lambda r_\mathrm{s}^2 \;,
\end{equation}
yielding the dimensionless effective potential
\begin{align}
\frac{V(x)}{G M/r_\mathrm{s}} =& -\frac{2 \tilde{j}^2 }{x^3}+\frac{2 \tilde{j}^2 +5}{4 x^2} +\frac{3 \tilde{\epsilon} -\frac{1}{2}}{x}  \nonumber \\ & +\left(\frac{\tilde{j}^2 \tilde{\Lambda}}{6}+\tilde{\epsilon} \right) +\frac{\tilde{\Lambda} x}{2} +\frac{1}{6} \tilde{\Lambda} x^2 \left(3 \tilde{\epsilon} -1\right) \;,
\label{eq:Veff}
\end{align}
where we take $\nu = 0$ without loss of generality. Since the maximum $\nu$ attainable is $1/4$, all signs in Eq.~\ref{eq:alphas} remain unchanged for the entire interval of admissible values for $\nu$. Hence, while $\nu=0$ only yields the effective potential for a massless test particle, the general case of $\nu\ne0$ is omitted for the sake of brevity and simplicity of equations shown.

\begin{figure}
    \centering
\includegraphics[width=0.44\textwidth]{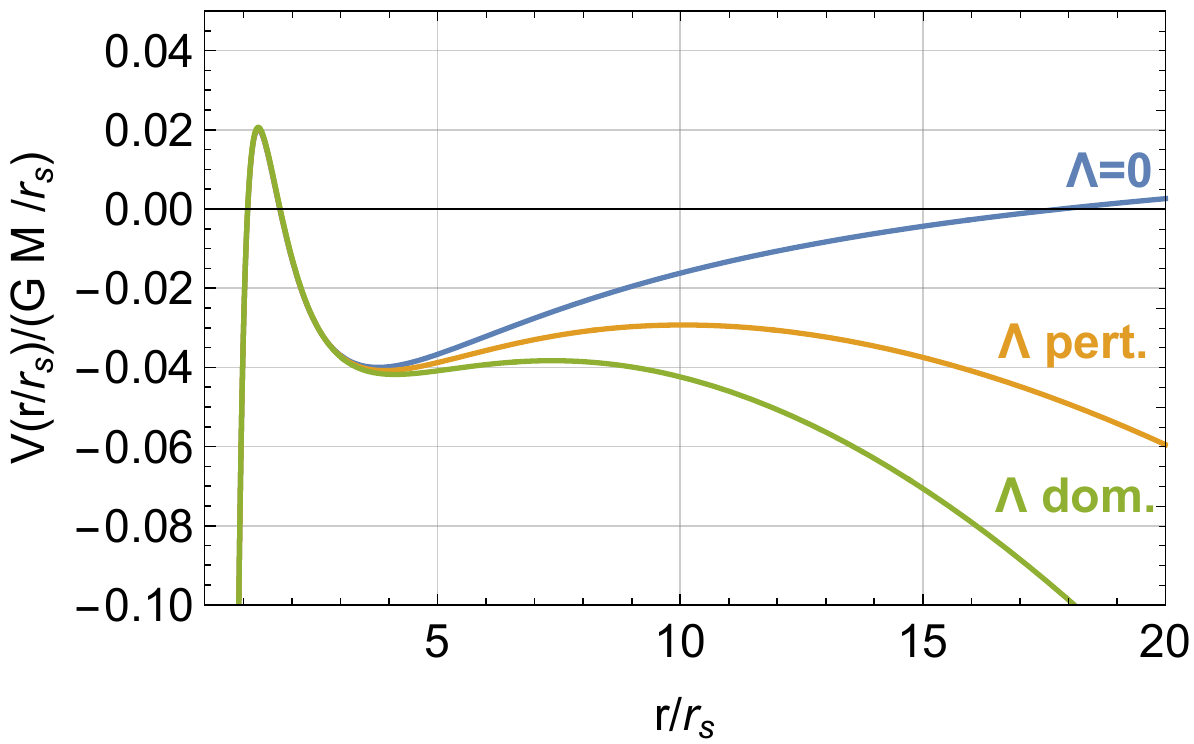}
\caption{Effective potential including post-Newtonian (PN) and cosmological-constant contributions scaled by the Newtonian potential at the Schwarzschild radius $r_\mathrm{s}=2GM/c^2$ versus the radius scaled by $r_\mathrm{s}$. For $\Lambda = 0$ (blue curve), stable orbits are reduced to the PN case. For small, perturbative $\Lambda$ (yellow curve), stable orbits can only become unbounded beyond a certain radius, where the impact of $\Lambda$ dominates. For large, dominating values of $\Lambda$ (green curve), there are only unbounded orbits.}
\label{fig:pot}
\end{figure}

Fig.~\ref{fig:pot} shows Eq.~(\ref{eq:Veff}) for different values of $\bar{\Lambda}$. We take $\bar{\Lambda} = 0$, $10^{-3}$, and $2 \cdot10^{-3}$ for the blue, yellow and green curves, respectively. For all of them, we choose $\epsilon/c^2 = -0.15$ and $\bar{j} = 2.65$ as arbitrary, representative example values. For small radii, the term $\sim r^{-3}$ is dominant. The first maximum in the effective potential which is caused by the interplay between the $r^{-3}$ and the $r^{-2}$ terms does not change much when changing $\Lambda$. At larger radii, the quadratic shape of the potential depends on the value of $\Lambda$. 

For $\bar{\Lambda} = 0$, stable orbits are reduced to the PN case. For small, perturbative $\bar{\Lambda} = 10^{-3}$, stable orbits can become unbounded beyond a certain radius, where the impact of $\Lambda$ dominates. For the value of $\bar{\Lambda} = 2\cdot10^{-3}$, there are only unbounded orbits.

\subsection{Equations of motion}
\label{sec:eom}

Similarly to \cite{Damour:1988mr}, we use the transformation on $r$
\begin{equation}
r = \bar{r} - \left(4 - \frac{3}{2} \nu\right) \frac{G M }{c^2 } \;,
\label{eq:r2rbar}
\end{equation}
that simplifies the potential for the $\dot{r}$ eliminating the $r^{-3}$ term. The equation of motion is reduced to
\begin{equation}
\frac{1}{2}\dot{\bar{r}}^2 = \epsilon_k + G\frac{m_k}{\bar{r}} - \frac{j_k^2}{2 \bar{r}^2} + k \bar{r} + \frac{1}{2} \Lambda_k \bar{r}^2 \;,
\label{eq:effectivePotenTrans}
    \end{equation}
with the abbreviations
\begin{eqnarray}
\epsilon_k &=& \epsilon \left[ 1 - \frac{3}{2} \left(1-3\nu \right)  \left(\frac{\epsilon}{c^2}- \frac{j^{2} \Lambda}{9\epsilon}\right) \right] \;,
\nonumber \\
m_k &=& M \left[ 1 - (6-7\nu) \frac{\epsilon}{c^2}\right] \;,
\nonumber \\
j_k^2 &=& j^2 \left[ 1 - 2(1-3\nu) \frac{\epsilon}{c^2} + 2(1-\nu) \left(\frac{G M}{j c}\right)^2\right] \;,
\nonumber \\
k &=& \frac{7 - 5 \nu}{3} G M \Lambda \;, \nonumber \\
\Lambda_k &=& \frac{1}{3}\Lambda  c^2 \left(1 + 3 \left(3\nu -1\right) \frac{\epsilon}{c^2}\right) \;. \nonumber 
\end{eqnarray}
Correspondingly, the transformation
\begin{equation}
r = \tilde{r} - \left(2-\nu\right) \frac{G M }{c^2}\;,
\end{equation}
simplifies the equation of motion for the true anomaly
\begin{equation}
\dot{\theta} = \left( 1 + \frac{\epsilon}{c^2}(3 \nu - 1) \right) \frac{j}{\tilde{r}^2} + \frac{ 3\nu - 1}{6}j \Lambda \;.
\label{eq:thetaTrans}
\end{equation}

\subsection{Closed Orbits}
\label{sec:closed}

In order to find the relation between the semi-major axis $a$ and the eccentricity $e$ to the variables in Section~\ref{sec:eom}, we use the condition where the $\dot{r}$ is zero:
\begin{equation}
\dot{r} \left[ a(1 \pm e )\right] = 0 \;.
\end{equation}
Imposing this condition on Eq.~(\ref{eq:effectivePotenTrans}) yields the relations for the energy and the angular momentum
\begin{eqnarray}
\epsilon &=& - \frac{G M}{2 a} \left[ 1 + \frac{2}{3} \left(1+e^2\right)\kappa + \beta \frac{\nu -7}{4} \right.
\nonumber \\
&&
\left. -\frac{2}{3} \beta \kappa \left(6 e^2+11 \nu -10\right) \right] \;,  
\nonumber \\
j^2 &=& G M a (1- e^2) \left[ 1 - \kappa\frac{1 - e^2}{3} + \beta\frac{4 - e^2 (\nu -2)}{ 1 -e^2} \right. \;,
\nonumber \\
& &\left. - \beta \kappa \frac{e^2 (6 \nu +1)-14 \nu +15}{6}\right] \;. 
\nonumber \\
\end{eqnarray}
For the limit $\kappa \rightarrow 0 $, we retrieve the same relations for the modified energy and angular momentum as for the pure PN case. Analogously, for $\beta \rightarrow 0$, we obtain the modified relations for the pure PC case as detailed in~\cite{Benisty:2023vbz}. For closed orbits, we parameterise the solution of $r$ as
\begin{equation}
\bar{r} = \bar{a}\left(1 - \bar{e} \cos\eta  \right) \;,
\end{equation}
in which $\eta$ is the mean anomaly angle and $\bar{a}$ and $\bar{e}$ are the transformed semi-major axis and eccentricity, respectively. Integrating Eq.~(\ref{eq:effectivePotenTrans}) under the assumption that $\beta $ and $\kappa$ are sufficiently small, we obtain
\begin{equation}
\frac{2 \pi}{T} \left(t - t_0\right) = \eta - e_t \sin \eta + \kappa \left( g_2 e^2 s_2 + g_3 e^3 s_3 \right) \;,  
\end{equation}
where $s_2 = \sin 2\eta$, $s_3 = \sin 3 \eta$, and analogously for higher orders not shown here. We also abbreviate
\begin{equation}
g_2 = \frac{5}{24} - \beta \frac{67}{48} \;, \quad g_3 = -\frac{1}{72} + \beta \frac{27-7 \nu}{144} \;.
\end{equation}
The modified temporal eccentricity and the period read
\begin{eqnarray}
\frac{e_t}{e} &=& 1 + \kappa \frac{16 - 7 e^2}{24} - \frac{\beta}{2}  (8-3 \nu ) \nonumber\\
& &
-\beta \kappa \frac{64 (\nu +1)-e^2 (133-9 \nu )}{48}\;, 
\nonumber\\
T &=& 2 \pi \sqrt{\frac{a^3}{G M}} \left[ 1 + \frac{\beta}{2} (9 - \nu) + \kappa \left(\frac{e^2}{4}+\frac{2}{3}\right) \right.
\nonumber\\
&&
\left. + \beta \kappa \frac{108 -56 \nu + e^2 (37 \nu -54)}{24} \right] \;.
\label{eq:Tfull}
\end{eqnarray}
The $g_2$ and $g_3$ couplings are due to the $\kappa $ order and inside their definitions is a modification of order of $\beta \kappa$.

Using Eq.~(\ref{eq:thetaTrans}), parameterising $\tilde{r}$ via the relation
\begin{equation}
\tilde{r} = \tilde{a} \left(1 - \tilde{e} \cos \nu \right) \;,
\end{equation}
we obtain the solution for the true anomaly
\begin{equation}
\frac{\theta}{2 (1+k)} = \text{tan}^{-1}\left[\sqrt{\frac{1 + e_{\phi}}{1 - e_{\phi}}} \tan\frac{\nu}{2} \right]- \frac{\kappa}{6} \left(3 \eta - e \sin \eta \right) 
\end{equation}
with $e_\phi = e  \left(1 + \frac{\nu}{2}\beta \right)$ and $k$ being the pericentre advance parameter introduced in Eq.~(\ref{eq:dom}). For a complete period running from $\nu = 0$ to $\nu = 2 \pi$, the excess angle is
\begin{equation}
\Delta \omega \equiv 2 \pi k = 6 \pi \frac{\beta}{1 - e^2}  +  \pi \kappa \sqrt{1-e^2} \;.
\label{eq:dom}
\end{equation}
Here $\omega$ is the argument of the pericentre. For $\kappa \rightarrow 0$, we get the same known precession term from the PN correction.

\subsection{Unbound orbits}
\label{sec:unbound}
For very large separations $r$, the $\Lambda$ terms, and thus cosmic repulsion, dominate in Eq.~(\ref{eq:eom}). To obtain an analytic solution for this regime, we consequently set up the radial equation of motion for an unbound orbit to be
\begin{equation}
\dot{\bar{r}}^2 = 2 \epsilon_k + 2 k \bar{r} + \Lambda_k \bar{r}^2 \;.
\label{eq:unbound_r}
\end{equation}
Taking the derivative of this equation with respect to $t$ and assuming that $\dot{\bar{r}} \ne 0$, we obtain
\begin{equation}
\ddot{\bar{r}} = k + \Lambda_k \bar{r} \;.
\end{equation}
The ansatz $\bar{r}(t) = A/\sqrt{\Lambda_k} \text{sinh}(\sqrt{\Lambda_k}t) + B \text{cosh}(\sqrt{\Lambda_k}t)$ with coefficients $A$ and $B$ determined from initial conditions is easily verified as a solution to Eq.~(\ref{eq:unbound_r}). 
To fix the boundary conditions that determine $A$ and $B$, let us assume that $\bar{r}(t=0)\equiv \bar{r}_0$ and $\dot{\bar{r}}(t=0)\equiv \bar{v}_0$. The solution to Eq.~(\ref{eq:unbound_r}) then reads:
\begin{equation}
\bar{r}(t) = \frac{\bar{v}_0}{\sqrt{\Lambda_k}} \text{sinh}(\sqrt{\Lambda_k}t) + \left(r_0 + \frac{k}{\Lambda_k}\right) \text{cosh}(\sqrt{\Lambda_k}t) - \frac{k}{\Lambda_k} 
\label{eq:urbar}
\end{equation}
with 
\begin{equation}
\bar{v}_0 = \pm \sqrt{2 \epsilon_k + 2 k \bar{r}_0 + \Lambda_k \bar{r}^2} \;.
\end{equation}
The plus or minus sign can be chosen depending on the initial direction of the motion.
Next, we use Eq.~(\ref{eq:r2rbar}) to transform Eq.~(\ref{eq:urbar}) back into the original coordinate system to find a physically reasonable initial position.
Requiring that $r(t=0) \ge 0$, we find that $\bar{r}_0 \ge (2-3/4\nu)~r_\mathrm{s}$. 
Depending on the energy of the reduced mass, $\epsilon_k$, the initial velocity is fixed as well, given the initial position. At the maximum $\nu=1/4$, we find that the initial condition for the position requires to start at a radius of at least $1.625~r_\mathrm{s}$, implying that the reduced mass is on the right hand side of the maximum effective potential belonging to the last stable orbit around the central mass as plotted for dominating $\Lambda$ in Fig.~\ref{fig:pot}.

Next, we solve the angular equation of motion for an unbound system. Similarly to neglecting the terms inversely proportional to a power-law of $r$, we neglect the first term on the right-hand side of Eq.~(\ref{eq:thetaTrans}) for an effective potential with dominating $\Lambda$. Then, $\dot{\theta}\propto j\Lambda$. Treating $j$ as a conserved and known quantity, the solution of the angular equation of motion is simply:
\begin{equation}
\theta(t) = \frac{3\nu - 1}{6} j \Lambda t \;, 
\end{equation}
assuming that the motion started at $t=0$ for consistency with the solution for $r(t)$. For the more general case of intermediate separations, the equations of motion have to be solved numerically. This is, for instance, done in \cite{Kim:2020gai} to study the interplay between gravitation and dark energy for the Virgo Cluster (see Section~\ref{sec:upper} for details).

Noting that an unbound orbit can also be understood as a scattering process with the reduced mass being scattered off the effective central potential, our approach can be connected to the one sketched in \cite{bib:Hertzberg}. The authors also use a de Sitter metric as an embedding for their scattering scenario. Then, they employ a timing argument that the scattering black hole needs to exist for at least twice the Hubble time to constrain $\Lambda$ of the de Sitter embedding based on the fact that the evaporation time of a black hole is linked to its mass and this, in turn, can be inferred from the scattering process. Most notably, their calculations arrive at a \emph{lower} bound on $\Lambda$. Hence, transferring this argument to our unbound system of the LG and the Virgo Cluster, it is obvious that follow-up studies for this system should also obtain a minimum $\Lambda$, necessary to turn this binary system into an unbound one (see also Section~\ref{sec:upper}).

\subsection{Osculating formalism}

Another way to approach the orbits is the osculating formalism ~\cite{poisson_will_2014}. It can be used to describe the orbits of a body moving in a gravitational field, taking into account all the perturbing forces acting on it at the moment under analysis. Keplerian orbits are assumed to be tangential to the true orbit at any moment. The equations that govern the orbital parameters in our case can be summarised to
\begin{equation*}
\frac{de}{df} = \frac{p^3}{G M} \left[\frac{s_1 \mathcal{R}}{\left(1 + e c_1 \right)^2}  + 2\frac{c_1  +e(1+c_1^2)}{\left(1+ e c_1\right)^3}\mathcal{S}\right] \;,
\end{equation*}  
\begin{equation*}
e \frac{d \omega}{d f} = \frac{p^2}{G M} \left[-\frac{c_1}{\left(1+ e c_1 \right)^2}\mathcal{R} + \frac{s_1(2+e c_1)}{\left(1+ e c_1  \right)^3}\mathcal{S}\right] \;,
\end{equation*}
\begin{equation*}
\frac{{dt}}{{df}} = \frac{\sqrt{\frac{p^{3}}{G M}}}{\left(1+e c_1\right)^{2}}
\left(1-\tfrac{p^{2}}{e G M} \left(\tfrac{c_1 \mathcal{R}}{\left(1+e c_1\right)^{2}}-\tfrac{\left(2+e c_1\right) s_1 \mathcal{S}}{\left(1+e c_1\right)^3}\right) \right) \;,
\end{equation*}
where the perturbing force is described by
\begin{equation}
\mathcal{R} = \kappa \, \frac{G M}{3a^2}\frac{1-e^2}{1 + e c_1} \;, \quad \mathcal{S} = 0 \;.
\end{equation}
To shorten the equations, we set $c_1 \equiv \cos{f}, s_1 \equiv \sin{f}$, in which $f$ is the true anomaly $\theta$ with initial value $\theta_0$ and $\omega$ is the angle between the ascending node and the pericentre.
The integration of the osculating equations gives two types of changes -- an oscillation with a period equal to the orbital period and a steady drift of the orbital elements that do not average out after few periods. For the secular drift, the integration over $[0,2\pi]$ yields
\begin{subequations}
\begin{equation}
\Delta e = 0 \,, \quad \Delta\omega =  \pi \kappa \sqrt{1-e^2} \;,
\end{equation}
\begin{equation}
\Delta T= 2 \pi  \sqrt{\frac{a^{3}}{G M}}\left[1 - \frac{5}{24} \left(3 e^2+4\right) \kappa  \right] \;.
\end{equation}
\end{subequations}
We use this formalism to confirm the results obtained for the closed orbits in Section~\ref{sec:closed}, see Eqs.~(\ref{eq:Tfull}) and (\ref{eq:dom}). As expected, the terms match the derivation in our unified framework. Here again, we see the explicit dependence on $\kappa$ relating the orbital elements to $T_\Lambda$.

\begin{figure*}
    \centering
\includegraphics[width=0.8\textwidth]{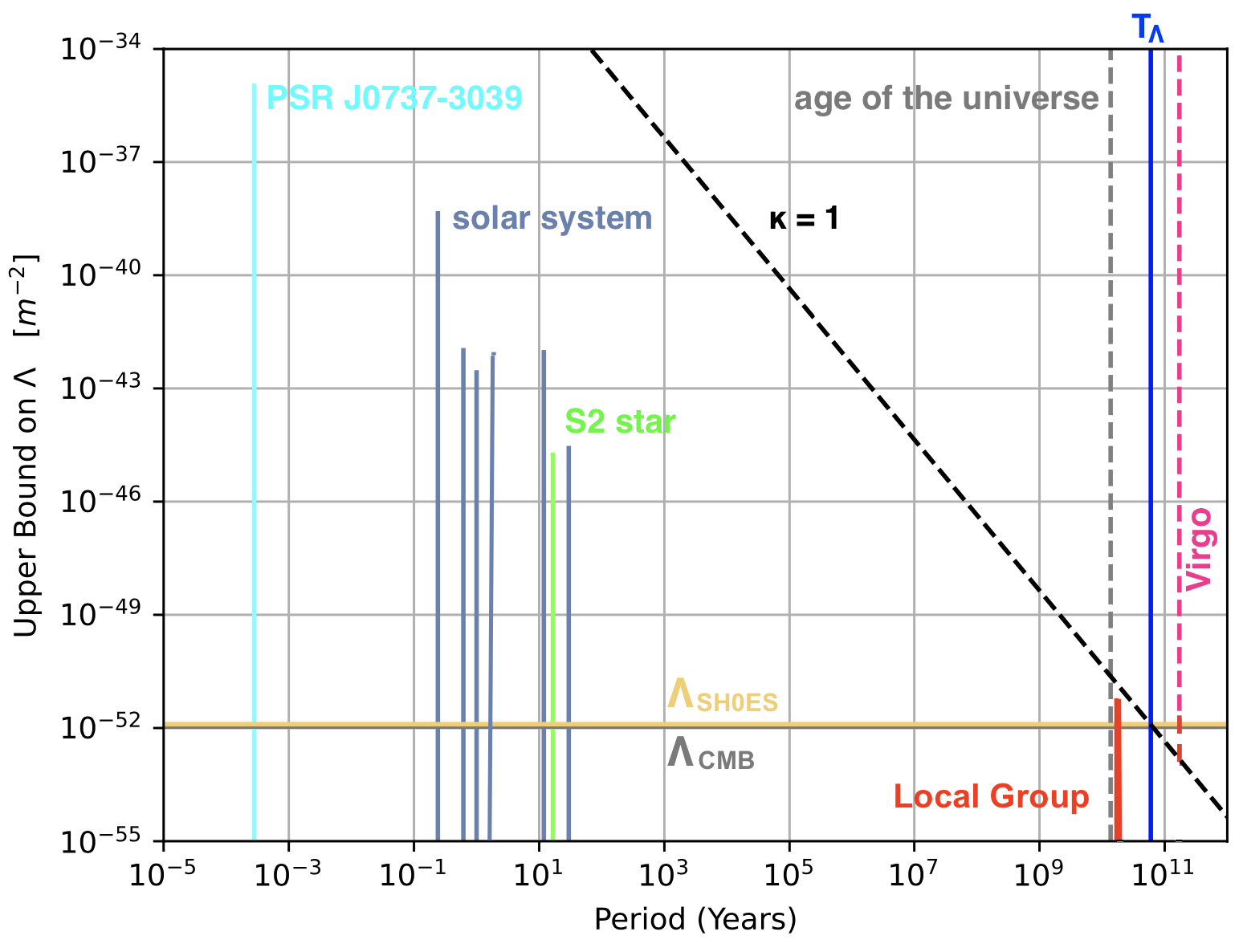}
\caption{Comparison between bounds on the cosmological constant for different systems versus the orbital period of these systems: planets in the solar system (dark blue), S2 star around Sgr A* (green), double pulsar PSR~J0737-3039A/B (azure), Local Group (red), and Virgo (pink). Each upper bound is the $1 $-$\sigma$ upper bound on $\Lambda$. For comparison, we put the estimated values of the cosmological constant from \cite{Planck:2018vyg} (grey) and SH0ES, \cite{Brout:2022vxf}, (yellow). For longer periods, the upper bound on $\Lambda$ decreases and proves that systems with orbital periods in the range of $T_\Lambda \approx 60~\text{Gyr}$ have to be used to constrain $\Lambda$ from binary systems tightly. The dashed black line is $\kappa = 1$ which represents $T_{\text{K}} = T_{\Lambda}$ and thus separates bound from unbound systems.}
\label{fig:upper}
\end{figure*}

\section{Bounds on the cosmological constant}
\label{sec:upper}

Fig~\ref{fig:periods} shows the relation between the orbital periods of different systems and $T_\Lambda$. The closer the orbital periods are to $T_\Lambda$, the tighter the constraints on the cosmological constant will be. However, as briefly noted in Sect.~\ref{sec:astro}, for binary systems fulfiling this criterion, like galaxies and galaxy clusters, we only have a snapshot of the orbital motion because possible observation times are much shorter than the orbital periods. 
For systems with much shorter orbital periods, like pulsars or stars orbiting around Sgr A*, we are able to track the entire orbit for several periods and constrain the corresponding precession. Yet, despite the increased precision, the lower orbital period will weaken the constraint that can be put on $\Lambda$ from such systems. 

In this section, we show the correlation between the precision and the bound on $\Lambda$ based on the observed periods of binary systems introduced in Section~\ref{sec:astro} taking their measurement precision into account.
To solve the equations of motions of Section~\ref{sec:eom} for the individual systems given the measurement precision of the velocities and distances to be inserted into the formalism, we use a $\chi^2$-approach with a flat prior of $\kappa \in [0,100]$, or $\Lambda \in [0,100] \times \omega_\text{K}^2/c^2$, where $\omega_\text{K}$ is the orbital angular frequency of the system. We use an affine-invariant nested sampler for the minimisation of the likelihoods as implemented in the open-source package \texttt{Polychord}~\cite{Handley:2015fda} and run it in the standard configuration, satisfying the needs of our optimisation problem.

\subsection{Solar System}
Upper bounds from the solar system were discussed in \cite{1999arse.conf..312W,Kagramanova:2006ax,Sereno:2006re,Iorio:2007ub,Liang:2014vma}. 
These different methods give different ranges of $\Lambda$. In our analysis, we use those planets whose orbital precession $\dot{\omega}_\text{m}$ has been measured: Mercury, Venus, Earth, Mars, Jupiter, and Saturn. We also take their planetary parameters from a NASA fact sheet\footnote{available at https://nssdc.gsfc.nasa.gov/planetary/factsheet/}.
All observations are inserted into 
\begin{equation}
\chi^2_\text{SS} = \left(\frac{\dot{\omega}(M,a,\Lambda) - \dot{\omega}_m}{\Delta\dot{\omega}_m}\right)^2
\end{equation}
to minimise the difference between the measured and modelled precession per the orbital period $\dot{\omega} = k/n$, which is a function of the planetary parameters and $\Lambda$. The posterior on $\Lambda$ for Mercury reads
\begin{equation}
\Lambda_{\text{Mercury}} = \left( 3.3633 \pm 2.227 \right) \times 10^{-39}\, \text{m}^{-2} \;, 
\end{equation} 
and the one for Saturn is
\begin{equation}
\Lambda_{\text{Saturn}} = \left( 1.104 \pm 0.9475 \right) \times 10^{-46}\, \text{m}^{-2} \;.
\end{equation} 
In Fig.~\ref{fig:upper}, we plot the upper bound for different planets over their orbital periods. Already for the solar system planets alone, the figure and the numbers show that the upper bound on $\Lambda$ is decreasing for increasing orbital period.

\subsection{Galactic Center}
Using the known precession of the S2 star orbiting Sgr~A*, we forecast a bound on $\Lambda$ with a $\chi^2$ similar to the one used for the solar system. The posterior on $\Lambda$ for S$_2$ yields:
\begin{equation}
\Lambda_{\text{S2}} = \left( 1.298 \pm 0.911\right) \cdot 10^{-46} \, \text{m}^{-2} \;.
\end{equation}
Comparing with the values from the solar system, this bound is of the order of the bounds from the outer planets, since the orbital period of S2 is about the same as Saturn.

\subsection{Pulsars}
PSR~J0737-3039A/B is the only known double pulsar with associated very high precision measurements. The system has been studied continuously using a number of radio telescopes, with improved data acquisition systems and better sensitivity, resulting in much improved timing precision over time. The latest measurement of PSR~J0737-3039A/B is published in~\cite{Kramer:2021jcw} and includes higher orders in the PN expansion to guarantee a high precision. The masses of the pulsar $m_p$ and its companion $m_c$ (which is another pulsar for the PSR~J0737-3039A/B event) have to be determined from the observables together with $\Lambda$. To do so, we calculate the power radiated in a de Sitter universe~\cite{Bonga:2017dlx}
\begin{equation}
P \sim \frac{32}{5 G} \left(G M_c n\right)^{10/3} \left(1 + \frac{5}{12}\kappa + \frac{\kappa^2}{36} \right) \;,
\end{equation}
where $M_c = (m_p m_c)^{3/5}/(m_p + m_c)^{1/5}$ is the chirp mass and $n$ is the orbital frequency. The $\chi^2$ for the double pulsar is thus
\begin{equation}
\chi^2_\text{PSR J0737} = \sum_{i = 1}^{3} \left(\frac{\xi(m_\text{p},m_\text{c},\Lambda) - \xi_\text{m}}{\delta \xi_\text{m}}\right)^2 \;,
\end{equation}
in which $\xi = (\dot{\omega}, \dot{P}, q)$ is the vector of the post-Keplerian parameters containing the measured values, and $\delta\xi$ represents the measurement uncertainties. The ratio $q \equiv m_\text{p}/m_\text{c}$ contains the mass of the pulsar and its companion. As priors for the post-Keplerian parameters, we use the Gaussian priors reported in the original papers. For the masses, we set a uniform prior of $[0,4] \times M_{\odot}$. The posterior for $\Lambda$ then yields
\begin{equation}
\Lambda_\text{PSR-J0737} = \left(4.64 \pm 3.74 \right) \times 10^{-36} \, \text{m}^{-2} \;.
\end{equation}
Despite the high precision of the pulsar observations, the upper bound is less tight than the one from the solar system or the S2 star due the shorter period of the system.

\subsection{Local Group}
The LG includes the Milky Way, the Andromeda galaxy, along with at least 78 other known members, most of which are dwarf galaxies. It has a total diameter of about 3~Mpc with the total mass of few $10^{12}~M_\odot$. The two largest members, the Andromeda Galaxy and the Milky Way, are both spiral galaxies of about $10^{12}~M_\odot$, with each hosting its own system of satellite galaxies. A simple model of the LG dynamics reduces the LG to these two galaxies as a two-body system.

Since the Keplerian period of the Milky Way and Andromeda ($17~\text{Gyr}$) is about $0.27 \, T_{\Lambda}$, \cite{Benisty:2023vbz} derives an upper bound on $\Lambda$ to be $5.44$ times the $\Lambda$-value obtained by Planck. The $\chi^2$ uses the mass of the LG and is taken from \cite{Benisty:2023vbz}:
\begin{equation}
\chi^2_\text{LG} =  \left(\frac{M (r,v_r, v_t, \Lambda) - M_{\rm m}}{\Delta M_{\rm m}} \right)^2,
\end{equation}
where $M (r,v_r, v_t, \Lambda)$ is the predicted mass as a function of the separation $r$ towards Andromeda, the tangential velocity $v_t$, the radial velocity $v_r$, and $\Lambda$. The, $M_{\rm m} \pm \Delta M_{\rm m}$ is the mass and its uncertainty inferred from the data as detailed in \cite{Benisty:2022ive}. 
It amounts to
\begin{equation}
M_{\rm m} \pm \Delta M_{\rm m} = \left(3.7 \pm 0.5 \right) \times 10^{12} M_{\odot} \;.
\end{equation}
Ref.~\cite{Benisty:2023vbz} uses a broad prior of $\Lambda = [0,10] \times \Lambda_{\rm CMB}$ and obtains the posterior on $\Lambda$, given the mean value of $\Lambda$ from the CMB denoted as $\bar{\Lambda}_\text{CMB}$
\begin{equation}
\Lambda = \left(3.07 \pm 2.37\right) \bar{\Lambda}_{\rm CMB} = \left(3.13 \pm 2.42\right) \times 10^{-52} \, \text{m}^{-2} \;.
\end{equation}
The posterior distribution includes the Planck value within less than one $\sigma$.

\subsection{Virgo Cluster}
The Keplerian orbital period for the binary system consisting of the LG and the Virgo Cluster is calculated from the total mass of the system being of the order of $M\approx 10^{15}~M_\odot$ and the semi-major axis of the orbit $a \approx 16~\text{Mpc}$. 
The local group with approximately $10^{12}~M_\odot$ only makes a minor contribution to the total mass. Besides this, we use the distance between the LG and Virgo as an approximation for the semi-major axis. 

For the obtained $T_\text{K}$, $\kappa >1$. Consequently, we conclude that this system is unbounded and derive a lower bound on $\Lambda$ from Eq.~(\ref{eq:kappa}) which yields
\begin{equation}
\Lambda_{\text{Virgo}} > 1.58 \times 10^{-53} \, \text{m}^{-2} \;.
\label{eq:LVirgo}
\end{equation}
While this estimate only considers the LG as a test particle in Virgo's extended, effective potential, the complex structure of the Virgo Cluster also allows to probe the potential for bounded orbits. Using 33 galaxies along the line of sight from our LG to the centre of the Virgo Cluster, \cite{Kim:2020gai} probe the effective potential to determine the radius at which gravity is balanced out by dark energy and $H_0$ from their infall model towards Virgo. The equations of motions are analogous to the ones in Section~\ref{sec:eom}, so that we can convert their constraint on $H_0$ to a constraint on $\Lambda$ using Eq.~(\ref{eq:LambdaExpansion}) to obtain
\begin{equation}
\Lambda_{\text{Virgo}} = \left(1.08 \pm 0.11 \right) \cdot 10^{-52}\, \text{m}^{-2} \;.
\end{equation}
This value is larger than the lower bound we calculated from Eq.~(\ref{eq:kappa}) and thus in agreement with our theoretical derivations.

\subsection{Planck versus SH0ES values}
Determining $\Lambda$ from the expansion rate of the universe (Eq.~\ref{eq:LambdaExpansion}) yields different values due the different values of the Hubble constant. The value from the Planck observations is
\begin{equation}
\Lambda_{\text{CMB}} = \left(1.097 \pm 0.02\right) \times 10^{-52} \, \text{m}^{-2} \;, 
\end{equation}
while the value from the SH0ES data yields
\begin{equation}
\Lambda_{\text{SH0ES}} = \left(1.26 \pm 0.05\right) \times 10^{-52} \, \text{m}^{-2} \;.
\end{equation}
Even though these two different values are in the range of $10^{-52}\, \text{m}^{-2}$, they have a $3$-$\sigma$ difference.

\subsection{Bounds versus orbital frequency}
Fig.~\ref{fig:upper} summarizes all upper bounds on $\Lambda$ for the different systems in terms of $1$-$\sigma$ upper bounds obtained by minimising the $\chi^2$ with respect to their measured orbital periods. Constraints on $\Lambda$ improve with increasing precision of the measurements. Therefore, we observe that the upper bounds on $\Lambda$ for the highly precise pulsar observations are far below the $\kappa=1$-line (dashed black line in Fig.~\ref{fig:upper}) which separates the bound orbits from the unbound ones. For the latter, $T_\text{K}=T_\Lambda$ by Eq.~(\ref{eq:kappa}), such that the attractive gravitation is balanced by repulsive dark energy (or zero-velocity surface as we discussed). Yet, even though the pulsar observations have a higher precision than the planets in the solar system or the S stars, the latter yield a stronger upper bound on $\Lambda$. The reason for this can be found in the fact that the periods for the planets in the solar systems are larger and the upper bound on $\Lambda$ comes closer to the $\kappa=1$-line, which is decreasing for increasing periods. Given that no bound on $\Lambda$ can cross this line without turning the bound system into an unbound one, it becomes clear that binary systems with increasing periods will yield tighter constraints on $\Lambda$.

Considering unbound systems like the LG and the Virgo Cluster, tightest lower bounds on $\Lambda$ can analogously be achieved close to the $\kappa=1$-line for systems with orbital periods close to $T_\Lambda$. Increasing the periods to much larger values, we arrive at regimes in which the cosmic expansion dominates and the orbiting bodies cannot be considered as a (perturbed) binary system anymore. 
In this regime, $\Lambda$ is rather constrained by supernovae or other large-scale observations. 

As an outlook, \cite{Iorio:2017auf} discusses the possibility for an upper bound on $\Lambda$ from future measurements of a pulsar in the vicinity of Sgr A*. According to our criterion of $\kappa$, the upper bound from this system will be optimal if the period of the pulsar is as long as possible and, at the same time, achieving the highest possible precision.

\section{Conclusion}
\label{sec:dis}
In this paper, we set up a general and unifying framework to solve the equations of motions for binary systems including both, the impact of dark energy, and the first post-Newtonian modification. We show that $T_{\Lambda} = 2\pi/c \sqrt{\Lambda} \approx 60~\text{Gyr}$ is a critical period for these systems. If the orbital period is much lower than $T_{\Lambda}$, dark energy only perturbs the binary motion and the gravitational attraction remains dominant. If the orbital period is much larger than $T_{\Lambda}$, the system is not bounded and the expansion of the universe, meaning the repulsive dark energy, is dominant. 

As detailed in \cite{Benisty:2023vbz}, the ratio between the Keplerian orbital period $T_\text{K}$ and $T_\Lambda$, turning $T_\Lambda$ into a critical time scale, arose from the Newtonian equations of motion when adding $\Lambda$.
Yet, all binary systems we consider can also be effectively described as Schwarzschild-like potentials embedded in a de Sitter spacetime. 
Doing so, see Sect.~\ref{sec:fundamentals}, we find that the same ratio between $T_\text{K}$ and $T_\Lambda$ naturally arises as the change in the Kretschmann scalar when embedding the Schwarzschild metric into the de Sitter spacetime to account for the impact of $\Lambda$ on the binary system, see Eq.~(\ref{eq:K2}). 
Thus, the ratio between the orbital period and the critical period is based on differential geometrical foundations beyond the purely Newtonian framework.

Hence, after the systematic analysis of binary systems in this unifying framework, we conclude that the solutions are characterised by two parameters: the dimensionless gravitational potential $\beta = \Phi/c^2$, selecting between Newtonian and post-Newtonian regime, and the ratio $\kappa = (T_{\text{K}}/T_{\Lambda})^2$, determining the influence of $\Lambda$ on the binary motion. 

Considering measurements from different binary systems having orbital periods from days to gigayears, we show that the upper bound on the cosmological constant decreases when the orbital period of the system increases. We derive upper bounds\footnote{defined as the 1-$\sigma$ upper bound of the posterior probability distribution around the mean $\Lambda$ value} on $\Lambda$ from the precession of solar system planets, the precession of the S2 star around the galactic center, and the precession of the double pulsar PSR~J0737-3039A/B. Despite its more precise measurements, the upper bound on $\Lambda$ from the double pulsar is less tight than the one obtained from S2 or Saturn, which yield the tightest constraints among these systems. This result is explained naturally in our generalised framework: As $\kappa=1$ separates bounded from unbounded systems, the increasing orbital periods of the S2 star and Saturn compared to the double pulsar require the upper bounds on $\Lambda$ to be below the $\kappa=1$-line in Fig.~\ref{fig:upper} in order not to turn bounded systems into unbounded ones. Consequently, the constraints on $\Lambda$ from bounded systems become tighter, the closer $T_\text{K}$ approaches $T_\Lambda$.  

The Local Group (LG) has the longest orbital period in bounded binary motion that we discuss in this paper. Since the orbital period of the LG is about $0.27~T_{\Lambda}$ ($\kappa\sim 10\%$), the upper bound on $\Lambda$ is expected to be tighter than the ones from the solar system and S2. Ref.~\cite{Benisty:2023vbz} found the upper bound on $\Lambda$ to be $5.44$ times larger than the Planck value, \cite{Planck:2018vyg}, which is an impressive result for an individual system in our local cosmic neighbourhood compared to the all-sky observations from the early universe. 
As Fig.~\ref{fig:periods} shows, further applications of our framework introduced here to galaxy rotation curves or even galaxy clusters should yield constraints on $\Lambda$ within the same precision range, which will be subject of further studies.

As $T_\text{K}<T_\Lambda$ for bounded orbits, they yield \emph{upper} bounds on $\Lambda$, while unbounded orbits with their $T_\text{K} > T_\Lambda$ yield \emph{lower} bounds on $\Lambda$. Strictly speaking, the latter case does not have ``orbits'' anymore, so one can also interpret it as a scattering of the test particle in the effective central potential, as is, for instance, considered in \cite{bib:Hertzberg}. 
One example for an unbound motion considered here is the system of the LG and the Virgo Cluster. Ref.~\cite{Kim:2020gai} probe the effective potential of the Virgo Cluster at increasing distances out to the LG to find the distance at which the gravitational attraction by the Virgo Cluster is balanced by the repulsion from $\Lambda$. Hence, this is the first binary system from which an upper and a lower bound on $\Lambda$ can be inferred and probably the one with the highest precision.
While $\Lambda=0$ could not have been excluded for the LG in \cite{Benisty:2023vbz} yet, current estimates for the Virgo Cluster, Eq.~(\ref{eq:LVirgo}), based on the results of \cite{Kim:2020gai} already challenge a vanishing $\Lambda$.

Trying to constrain dark energy models beyond a constant $\Lambda$ is more complicated, as binary systems with the right order of magnitude of orbital periods, $T_\text{K} \approx T_\Lambda$, are usually subject to lower observational precision (to be performed at larger distances from our position). The evaluation also involves more model dependencies as the increased orbital periods only allow us to observe a tiny fraction of the orbit. 
Additionally increasing the number of degrees of freedom to be constrained will lead to weaker confidence bounds due to degeneracies, so that we first aim at constraining $\Lambda$ as tightly as possible in follow-up studies before turning to more complicated models.

\begin{acknowledgements}
We thank to Avi-Loeb, Anne-Cristine Davis and Moshe Chaichian for stimulated discussions and comments. This work was done in a Short Term Scientific Missions (STSM) in Bulgaria, funded by the COST Action CA21136 "Addressing observational tensions in cosmology with systematics and fundamental physics (CosmoVerse)''. D.B. thanks the Carl-Wilhelm Fueck Stiftung and the Margarethe und Herbert Puschmann Stiftung. D.S. acknowledges the support of Bulgarian National Science Fund No. KP-06-N58/5. We have received partial support from European COST actions CA15117 and CA18108.
\end{acknowledgements}
\bibliographystyle{aa}
%
\bibliography{ref}




\end{document}